\documentclass[twocolumn]{aastex631}

\shorttitle{Modeling outbursts of EX~Draconis}
\shortauthors{Schlindwein \& Baptista}
\graphicspath{{./}{figures/}}
\begin{document}

\title{Mass-transfer Outburts reborn: \\
  Modeling the light curve of the dwarf nova EX~Draconis}

\correspondingauthor{Wagner Schlindwein}
\email{wagner.schlindwein@astro.ufsc.br}

\author[0000-0002-7095-4147]{Wagner Schlindwein}
\affiliation{Departamento de Física, Universidade Federal de Santa Catarina, Campus Trindade, Florianópolis-SC, Brazil}
\affiliation{Instituto Nacional de Pesquisas Espaciais, Avenida dos Astronautas, 1758, São José dos Campos-SP, Brazil}

\author[0000-0001-5755-7000]{Raymundo Baptista}
\affiliation{Departamento de Física, Universidade Federal de Santa Catarina, Campus Trindade, Florianópolis-SC, Brazil}

%% Note that the \and command from previous versions of AASTeX is now
%% depreciated in this version as it is no longer necessary. AASTeX 
%% automatically takes care of all commas and "and"s between authors names.

%% AASTeX 6.31 has the new \collaboration and \nocollaboration commands to
%% provide the collaboration status of a group of authors. These commands 
%% can be used either before or after the list of corresponding authors. The
%% argument for \collaboration is the collaboration identifier. Authors are
%% encouraged to surround collaboration identifiers with ()s. The 
%% \nocollaboration command takes no argument and exists to indicate that
%% the nearby authors are not part of surrounding collaborations.

%% Mark off the abstract in the ``abstract'' environment. 
\begin{abstract}

EX~Draconis is an eclipsing dwarf nova that shows outbursts with moderate amplitude ($\simeq 2$\,mag) and a recurrence timescale of $\simeq 20$-30\,d.
Dwarf novae outbursts are explained in terms of either a thermal-viscous instability in the disc or an instability in the mass transfer rate of the donor star (MTIM).
We developed simulations of the response of accretion discs to events of enhanced mass transfer, in the context of the MTIM, and applied them to model the light curve and variations in the radius of the EX~Dra disc throughout the outburst.
We obtain the first modeling of a dwarf nova outburst by using $\chi^2$ to select, from a grid of simulations, the best-fit parameters to the observed EX~Dra outbursts. The observed time evolution of the system brightness and the changes in the radius of the outer disc along the outburst cycle are satisfactorily reproduced by a model of the response of an accretion disc with a viscosity parameter $\alpha = 4.0$ and a quiescent mass transfer rate $\dot{M}_2 (\textrm{quiescence}) = 4.0 \times 10^{16}$\,g/s to an event of width $\Delta t = 6.0 \times 10^5$\,s ($\sim 7$\,d) where the mass-transfer rate increases to $\dot{M}_2 (\textrm{outburst}) = 1.5 \times 10^{18}$\,g/s.

\end{abstract}

%% Keywords should appear after the \end{abstract} command. 
%% The AAS Journals now uses Unified Astronomy Thesaurus concepts:
%% https://astrothesaurus.org
%% You will be asked to selected these concepts during the submission process
%% but this old "keyword" functionality is maintained in case authors want
%% to include these concepts in their preprints.

%% From the front matter, we move on to the body of the paper.
%% Sections are demarcated by \section and \subsection, respectively.
%% Observe the use of the LaTeX \label
%% command after the \subsection to give a symbolic KEY to the
%% subsection for cross-referencing in a \ref command.
%% You can use LaTeX's \ref and \label commands to keep track of
%% cross-references to sections, equations, tables, and figures.
%% That way, if you change the order of any elements, LaTeX will
%% automatically renumber them.
%%
%% We recommend that authors also use the natbib \citep
%% and \citet commands to identify citations.  The citations are
%% tied to the reference list via symbolic KEYs. The KEY corresponds
%% to the KEY in the \bibitem in the reference list below. 

\section{Introduction}

In dwarf novae, a late-type star (the secondary) transfers matter to a white dwarf companion through an accretion disc. Dwarf novae show recurrent outbursts at days-months timescales, in which the accretion disc increases in brightness by factors 20-100. Two models compete to explain the causes of these outbursts. The disc instability model \citep[DIM,][]{Cannizzo1993,Lasota2001,Hameury2020} explains the outbursts in terms of a thermal-viscous instability in the disc that causes it to cyclically transition between a cold and low viscosity state (quiescence) and a hot and high viscosity state (outburst). On the other hand, the mass transfer instability model \citep[MTIM,][]{Bath1972,Bath1975,BathPringle1981} explains the outbursts in terms of the response of a disc with constant (and high) viscosity to sudden increases in the mass transfer rate from the secondary.

Interest in the observational testing of the two models declined from the 1990s onward as a result of the widespread acceptance of DIM as the correct explanation. Two arguments were crucial in establishing the dominance of DIM, both based on the hypothesis that the mass transferred from the secondary is deposited only and always at the outer edge of the disc, in the position of the bright spot. On the basis of this hypothesis, it can be predicted that (i) a sudden increase in the mass transfer rate would inevitably lead to an increase in bright spot luminosity, and (ii) the MTIM mechanism could only produce outside-in outbursts because the excess matter would always be deposited at the outer edge of the disc. The existence of inside-out outbursts and the lack of compelling evidence for the increase in bright spot luminosity at the beginning of the outbursts were taken as arguments against MTIM \citep[and references therein]{Warner1995}.

However, numerical simulations of accretion discs show that when the gas stream is denser than the disc material, the stream ``penetrates'' the disc and allows matter to be deposited in its inner regions, giving rise to inside-out outbursts while leaving no enhanced bright spot emission footprint at outburst onset \citep{Bisikaloet1998a,Bisikaloet1998b,Makitaet2000}. The reader is referred to \citet{BaptistaSchlindwein2022} for a recent, comprehensive discussion on the stream-disc interaction, in particular on the possibility of stream penetration when the gas stream is denser than the disc gas, and for the definition of the penetration radius, $R_p$, as the radius where the midplane stream density equals that of the accretion disc. Additionally, strong observational support in favor of MTIM has emerged in recent years from a series of experiments monitoring the evolution of the brightness distribution of dwarf novae accretion discs throughout outbursts \citep{BaptistaCatalan2001,Baptistaet2007,Baptista2012,Baptistaet2016}, as well as the inference of high values for viscosity in discs of quiescent dwarf novae -- inconsistent with DIM predictions \citep{BaptistaBortoletto2004,Baptista2012,Baptistaet2016}. These results made it clear that there is a significant group of dwarf novae, the outbursts of which are incompatible with DIM and are presumably produced by the MTIM mechanism, leading to the suggestion that the two mechanisms coexist, probably in distinct subgroups of dwarf novae \citep{Baptista2012}.

MTIM assumes that outbursts are caused by the response of a disc with constant viscosity to a sudden increase in the mass transfer rate ($\dot{M}_2$). The value of the viscosity parameter $\alpha$\footnote{The solution is known as the steady $\alpha$-disc model and is obtained by adopting the prescription of \citet{ShakuraSunyaev1973}, $\nu = \alpha c_s H$, where $\alpha$ is a dimensionless parameter, $c_s$ is the speed of sound, and $H$ is the vertical scaleheight.}
 % One of the ways to interpret this expression is to assume that the
 % viscosity is generated by swirls of diameter $H$ and rotational speed
  % $\alpha c_s$.
is obtained from the observed timescale for the decay of outburts in dwarf novae, $\alpha\sim 1-3$ \citep{MantleBath1983}. The cause of the sudden increase in $\dot{M}_2$ may be associated with the convective nature of the secondary.
\citet{Paczynski1965} was the first to point out that convective low-mass stars, like the secondaries typical of cataclysmic variables (CVs), are potentially unstable. Shortly afterwards, \citet{Bath1972} showed that the region of instability in the envelope becomes much more extensive with the application of boundary conditions similar to Roche equipotentials, reaching the conclusion that stars that would normally be stable in the face of dynamical instabilities can become unstable when filling their Roche lobes in a semi-detached binary. An alternative explanation for the sudden increases in $\dot{M}_2$ involves the interaction of starspots with the inner Lagrangian point ($L_1$). \citet{LivioPringle1994} and \textcite{KingCannizzo1998} argued that the passage of starspots in front of the $L_1$ point can significantly reduce the mass transfer rate in CVs. \citet{HameuryLasota2014} reversed the argument to propose that the sudden variations in $\dot{M}_2$ required by MTIM can occur at the ocasions where there are no starspots transiting $L_1$.

In the 1980s, numerical simulations were carried out, solving the gas diffusion equation in the disc under the action of viscosity $\nu$, for the response of a viscous disc ($\alpha= 0.1-1$) to a short duration, sudden increase in mass transfer rate, but limited to the cases where the viscosity parameter $\alpha$ is constant both in time and with distance from the star, and only for a mass deposit restricted to the outer edge of the disc \citep{Pringle1981,BathPringle1981}.

\citet{Bathet1983} studied the effects of gas stream penetration during the outburst, obtaining results that can explain the different types of outbursts.  Outside-in outbursts occur when the material from the stream is deposited in the outermost regions of the disc, while inside-out outbursts occur when the stream penetrates the disc, depositing the material in more internal regions close to the circularization radius ($R_c$). \citet{Bathet1986} found that the shape of the outburst is quite sensitive to the shape of the assumed mass-transfer event, with the effect being more dramatic at outburst rise. \textcite{LivioVerbunt1988} simulated the response of a viscous disc to a sudden increase in the mass transfer rate, finding that the disc first shrinks (due to the sudden addition of gas with low angular momentum at the outer edge of the disc) and then expands.

\citet{IchikawaOsaki1992} performed dwarf nova outburst simulations within both the DIM and MTIM frameworks. However, as the focus of the work was on DIM, their MTIM results were limited both is scope and range of possibilities. Progress in the interpretation of observations of dwarf nova outbursts within the MTIM framework demands a better understanding of the response of a viscous disc to an event of enhanced mass-transfer, with the exploration of the entire parameter space of the model.

Here we report the results of more extensive simulations of accretion disc outbursts in the MTIM framework, and their application to the modeling of outburst observations of the dwarf nova EX~Draconis. The numerical methods and performed simulations are described in Sect.~\ref{sec:modelo}, while the application to EX~Draconis is reported in Sect.~\ref{sec:aplicacao}. The results are discussed in Sect.~\ref{sec:discursao} and summarized in Sect.~\ref{sec:conclusao}.

\section{Methods of Numerical Calculations} \label{sec:modelo}

We developed a computational code to simulate the response of accretion discs to events of enhanced mass transfer, in the context of the MTIM. We postulate the existence of variations in mass transfer rate, with no attempt to explain their nature.

In our simulations, we adopted the usual procedure of describing the behavior of the disc only in the radial direction. This is justified by the assumption that asymmetries in the distribution of mass in the azimuthal and vertical directions of the disc are eliminated faster than radial asymmetries, given the typical azimuthal (Keplerian) velocities of $\sim 1000$\,km/s, the local sound speed of $\sim 10$\,km/s (which determines the propagation of vertical disturbances in the disc), and the radial velocity of viscous flow $\sim 1$\,km/s.

The program is written in C programming language and follows the steps of \citet{IchikawaOsaki1992}, for which the reader is referred to for more details on the calculation of the time evolution of the accretion disc. It should be noted that the viscosity expression ($\nu$) assumed by \citet{IchikawaOsaki1992} includes an additional 2/3 factor for the $\alpha$-prescription of \citet{ShakuraSunyaev1973}. The code assumes a nonmagnetic primary, and initially starts from an already formed, steady-state thin disc (set of Eqs.~\ref{eq:conjunto_disco} in Appendix~\ref{sec:apendice_disco}). This last assumption helps reduce the required computational time.

\begin{figure}[!htp]
\includegraphics[width=1.0\columnwidth]{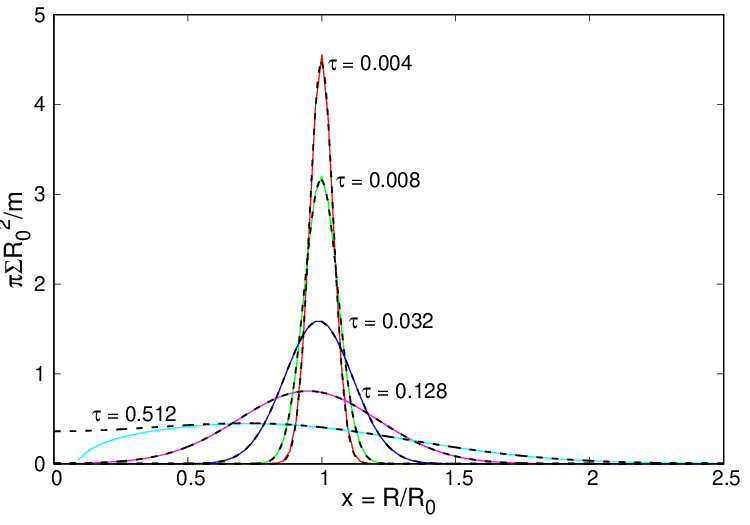}
\caption{Test of the viscosity efficiency in our simulations. A ring of mass $m$ placed in a Keplerian orbit at $R = R_0$ spreads under the action of viscous torques. The surface density $\Sigma$ is shown as a function of $x = R / R_0$ and the dimensionless time variable $\tau = 12 \nu t R_0^{-2}$, with $\nu = const$. The solid curves represent the solutions obtained through our simulations, while the dashed curves are the analytical solutions obtained by \citet{Pringle1981}. \label{fig:pringle1981}}
\end{figure}

The developed program was validated by testing its ability to reproduce the results obtained by \citet{Pringle1981} and \citet{IchikawaOsaki1992}. To reproduce the analytical solutions of \citet{Pringle1981}, we adopted the binary parameters of the dwarf nova V4140~Sgr \citep{BorgesBaptista2005}, a fixed viscosity of $\nu = 2.0 \times 10^{14}$\,cm$^2$\,s$^{-1}$, and sliced the disc into 100 concentric annular rings that extend up to 0.8 of the inner Lagrangian radius ($R_{L_1}$). Initially we deposit $10^{21}$\,g of matter in a ring at a radius of $9.55 \times 10^9$\,cm and let the system evolve over time. The results of this test are shown in Fig.~\ref{fig:pringle1981}. The simulations reproduce well the analytical results obtained by \citet{Pringle1981}. The only difference occurs for $x \rightarrow 0$ at $\tau=0.512$ (cyan curve in Fig.~\ref{fig:pringle1981}), where the numerical solution differs from the analytical one because the internal radius of the disc coincides with the radius of the white dwarf in the numerical simulation while the disc extends to the origin in the analytical solution of \citet{Pringle1981}.

\begin{figure*}[!htp]
\includegraphics[width=1.0\textwidth]{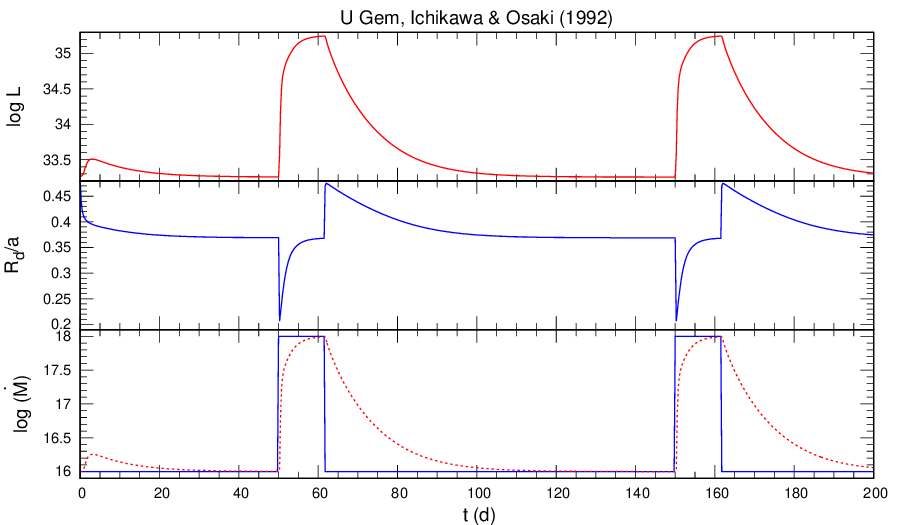}
\caption{Time evolution of the accretion disc of U~Gem obtained from our simulations for an event of enhanced mass of $10^6$\,s. \textit{Top panel:} The luminosity of the disc as a function of time (in units of days). \textit{Middle panel:} The radius of the disc in units of orbital separation $a$. \textit{Bottom panel:} The mass transfer rate (solid line) and the mass accretion rate over the primary (dotted line). \label{fig:teste_ichikawa_osaki1992}}
\end{figure*}

The next step in the validation process consists of reproducing the MTIM simulations by \citet{IchikawaOsaki1992}. The comparison is shown for their longer event duration of $10^6$\,s as it is easier to visualize the morphology of the results. We adopt the same binary parameters of the U~Gem dwarf nova assumed by these authors in our simulations with $\alpha=2/3$ \citep[which corresponds to the $\alpha=1$ of][]{IchikawaOsaki1992}. The disc was sliced into 100 rings of equal radial width extending up to $0.8\,R_{L_1}$. We adopted a time step of $\Delta t = 0.2$\,s. The simulation starts from a surface density radial distribution ($\Sigma (R)$) equals to that of a steady thin disc and with an outer disc radius $R_d = 0.8 \, R_{L_1}$. We let the disc evolve for 50\,d and then trigger the event of enhanced mass, and repeat this process after another 100\,d. The results are shown in Fig.~\ref{fig:teste_ichikawa_osaki1992}. The mass and angular momentum are conserved with relative accuracy better than $10^{-15}$ and $10^{-10}$, respectively. The results are in very good agreement with those of \citet{IchikawaOsaki1992} and can be directly compared to their Fig.~2.

\begin{figure}[!htp]
\centering
\includegraphics[width=1.0\columnwidth]{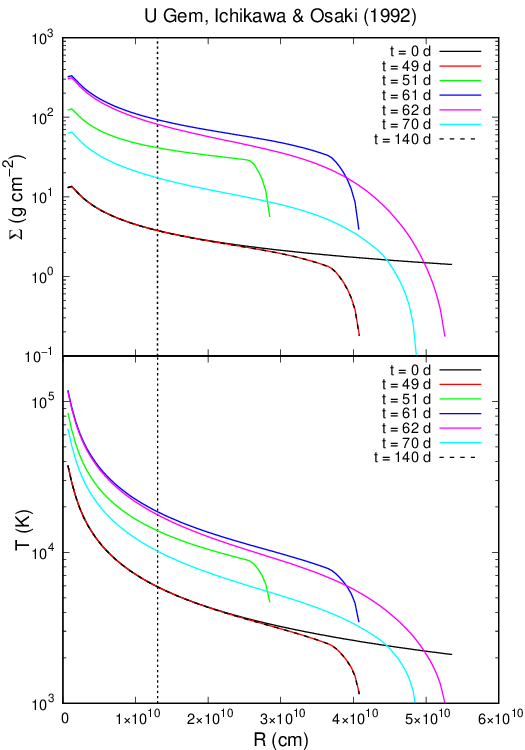}
\caption{Time evolution of surface density (top panel) and effective temperature (bottom panel) obtained from our simulations. The time corresponding to each curve is indicated in the figure legend. The dotted vertical line indicates the position of the circularization radius. \label{fig:teste_ichikawa_osaki1992_2}}
\end{figure}

Fig.~\ref{fig:teste_ichikawa_osaki1992_2} additionally shows the time evolution of the $\Sigma(R)$ and $T_\mathrm{eff}(R)$ distributions throughout the outburst cycle; the times indicated in the body of the figure are relative to Fig.~\ref{fig:teste_ichikawa_osaki1992}.

Analyzing the time evolution of $\Sigma$ and $T_\mathrm{eff}$, we observe that initially the tidal effect acts to truncate the distributions in the outer radius of the disc, drastically reducing the initial values of $\Sigma$ and $T_\mathrm{eff} $ in the outermost regions; this adjustment results in a small decline in luminosity lasting $\sim 30$\,d. The disc shrinks significantly and rapidly at outburst onset ($t=51$\,d) but gradually returns to its quiescent radius during the mass-transfer event ($t=61$\,d); at the end of the mass-transfer event, the disc expands rapidly reaching its largest radius ($t=62$\,d), thereafter slowly declining back to its quiescent radius -- in agreement with the results of \citet{LivioVerbunt1988}; 8\,d after maximum the disc is still much hotter and larger than in quiescence ($t=70$\,d). During the mass-transfer event (i.e., outburst rise), the $\Sigma$ and   $T_\mathrm{eff}$ distributions become flatter than the steady-state model in the outer disc regions; this is due to the larger amount of matter being deposited at the edge of the disc (at $R \geq 0.9\,R_d$). A viscous time must pass for the system to reach a steady state, and thus redistribute the additional mass throughout the disc. During the decay phase the disc cools and shrinks, in a manner that can be reasonably well described as a sequence of steady discs with progressively lower accretion rates.

Fig.~\ref{fig:teste_ichikawa_osaki1992_2} illustrates the reasons that led us to adopt different prescriptions for the width of the rings in the internal and external disc regions. With the lineary spaced rings of Fig.~\ref{fig:teste_ichikawa_osaki1992_2}, the $T_\mathrm{eff}(R)$ distribution is subsampled in the inner disc regions. As these are the regions that contribute most of the disc's the integrated disc luminosity, this introduces systematic errors in the calculation of luminosity and its variation throughout the outburst. Because of this, we decided to adopt logarithmic width rings in the $R<R_c$ regions. On the other hand, adopting logarithmic width rings across the entire length of the disc causes the rings in the outer regions to be very wide, compromising the correct assessment of the outer disc radius and its time variation. Therefore, we decided to adopt rings of logarithmic width in the inner disc regions ($R<R_c$) and rings of linear width in the outer disc regions (R$>R_c$). Therefore, the model has $N/2$ rings for $R<R_c$ and additional $N/2$ rings for $R>R_c$, where $N$ is the total number of rings in the disc.

With the basis of the simulations validated, we expanded the horizon of the MTIM simulations \citep{Schlindwein2021}.

\subsection{Variable $\alpha$} \label{sec:alpha_variavel}

First, we studied the effects of assuming a radial dependence for the viscosity parameter $\alpha$ on the outburst. As remarked by \citet{ShakuraSunyaev1973}, the introduction of the parameter $\alpha$ allows for an algebraic treatment of the equations that describe the physics of an accretion disc, transferring all ignorance of the mechanism responsible for viscosity to $\alpha$. There is no particular physical reason for $\alpha$ to be the same at all radii. We tested the hypothesis $\alpha= \alpha(R)$ using the following prescription,

\begin{equation}
\alpha = \alpha_0 \, \left( \frac{R}{R_c} \right)^n \,\,\, \longrightarrow \,\,\, \nu = \alpha_0 \, \left( \frac{R}{R_c} \right)^n \, c_s \, H,
\end{equation}

\noindent where $\alpha_0$ is a constant and $n$ is a number that represents the behavior of $\alpha$ in the radial direction. Since the outer disc radius varies along the outburst, we chose $R_c$ as the reference radius in the viscosity prescription. For $n>0$, viscosity is higher in the outer parts of the disc, whereas for $n<0$, viscosity is higher near the white dwarf; when $n=0$, we fall into the classical case where $\alpha$ is independent of radius. With this prescription, the equations for the radial structure of a steady thin disc (set of Eqs.~\ref{eq:conjunto_disco}) are modified by transforming $\alpha \rightarrow \alpha_0 (R/R_c)^n$. This transformation also applies to the equations used in our simulations.

We conducted simulations with the same configuration as that used in Fig.~\ref{fig:teste_ichikawa_osaki1992} to allow easy visualization of the resulting effects. The case of decreasing $\alpha$ with radius results in longer outburst decay for both the disc luminosity and outer radius in comparison to the case with constant $\alpha$. The result is the opposite for the case of increasing $\alpha$ with radius. This effect is easy to explain: the outer disc regions dominate the outburst decay timescale because they are farther away from the disc center and it takes a longer time for the disc gas there to flow onto the central object. Therefore, if these regions have a lower/higher viscosity, the outburst decay will correspondingly be longer/shorter. The interaction between the gas stream and the outer disc regions and the tidal effects of the secondary star (which preferentially affect the outer parts of the disc) are additional sources of energy dissipation whose effect on the disc structure is similar to an increase in the viscosity parameter with radius. In other words, these effects contribute to shorten the decay of the outburst and minimize the exponential tail that arises in MTIM simulations without their inclusion.

\begin{figure*}[!htp]
\includegraphics[width=1.0\textwidth]{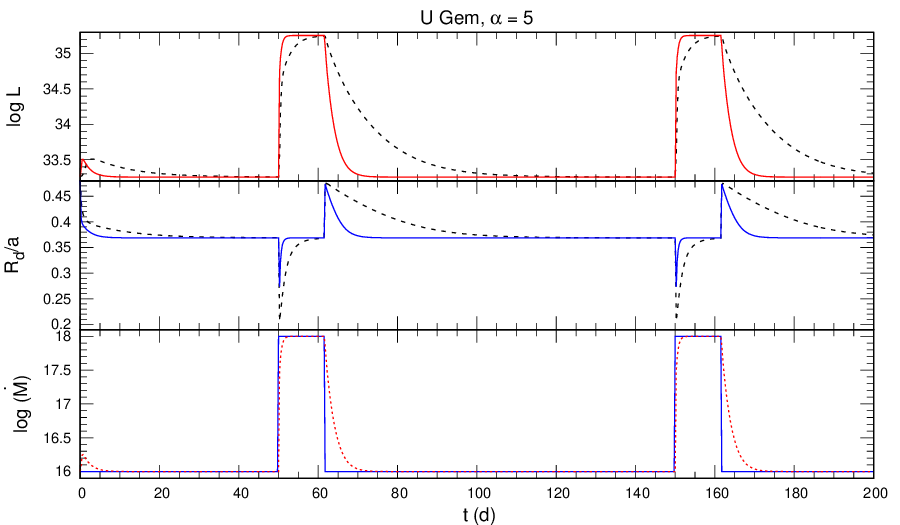}
\caption{Time evolution of the accretion disc with $\alpha = 5$. Analogous to Fig.~\ref{fig:teste_ichikawa_osaki1992}, but the dashed lines in the two upper panels are the results obtained in Sect.~\ref{sec:modelo}. \label{fig:alpha_5}}
\end{figure*}

Finally, we performed simulations for values of $\alpha$ above unity. Fig.~\ref{fig:alpha_5} shows the results of a simulation with $\alpha = 5$. As can be seen, both the rise time and especially the decay time of the outburst are significantly reduced, making the MTIM simulations more akin to outburst observations. This indicates that the MTIM model is capable of adequately reproducing dwarf nova outburst shapes by adopting values $\alpha \gtrsim 1$, in line with the $\alpha$ range inferred from dwarf nova outburst decline timescales by \citet{MantleBath1983}. It is worth emphasizing \citep{BaptistaSchlindwein2022} that the value of $\alpha$ inferred from the decline time of an outburst depends on the assumed model: In DIM, the decline of the outburst is due to the propagation of a cooling wave toward the center of the disc with velocity $R/t= v_\mathrm{DIM} \sim \alpha c_s$, where $R$ is the radius of the disc and $t$ is the observed decay time with $c_s\simeq 10$\,km/s. Velocities $v_\mathrm{DIM} \sim 1$\,km/s imply values $\alpha_\mathrm{DIM} \sim 0.1$. On the other hand, in MTIM the decline is due to the viscous flow of the mass in the disc with velocity $R/t = v_R \sim 12 \alpha c_s H/R$, where $v_R$ ($\simeq 1$\,km/ s) is the radial drift velocity. Therefore, for the same observed ratio $R/t$ we have the following,

\begin{eqnarray}
\frac{R}{t} \sim \alpha_\mathrm{DIM} \, c_s & \sim & 12 \, \alpha_\mathrm{MTIM} \, c_s \, \frac{H}{R} \nonumber \\
\alpha_\mathrm{MTIM} & \sim & 10 \, \alpha_\mathrm{DIM}, \label{eq:alpha_DIM_MTIM}
\end{eqnarray}

\noindent with $H/R \sim 10^{-2}$ (thin disc approximation). Thus, for a given outburst, the $\alpha$ value inferred with the MTIM assumption is an order of magnitude higher than with the DIM assumption.

\subsection{Shape of the enhanced mass event}

We tested the influence of the shape of the enhanced mass event on the morphology of the outburst. The box-shaped event used by \citet{IchikawaOsaki1992} is useful for highlighting the viscous response of the disc to a sudden increase/decrease in $\dot{M}$, but it is unrealistic. In real situations, a smoother transition is expected to occur between the quiescent and outburst $\dot{M}_2$ value.

Some formats were tested, and the one that produced the most interesting characteristics to describe the outbursts of dwarf novae was a modification of a Gaussian event with the following form,

\begin{equation}
\dot{M}_2 = \dot{M}_2^i + (\dot{M}_2^p-\dot{M}_2^i) \exp \left[ -\frac{1}{2} \left( \frac{t-t_p}{\Delta t_p/ \left( 2 \sqrt{2 \ln 2} \right)} \right)^4 \right],
\label{eq:pulso}
\end{equation}

\noindent where $\dot{M}_2^i$, $\dot{M}_2^p$, $t_p$ and $\Delta t_p$ are the quiescent and outburst mass transfer rates, the event center instant and its full-width-at-half-maximum, respectively. We will refer to this format as a quadri-Gaussian event.

\begin{figure*}[!htp]
\includegraphics[width=1.0\textwidth]{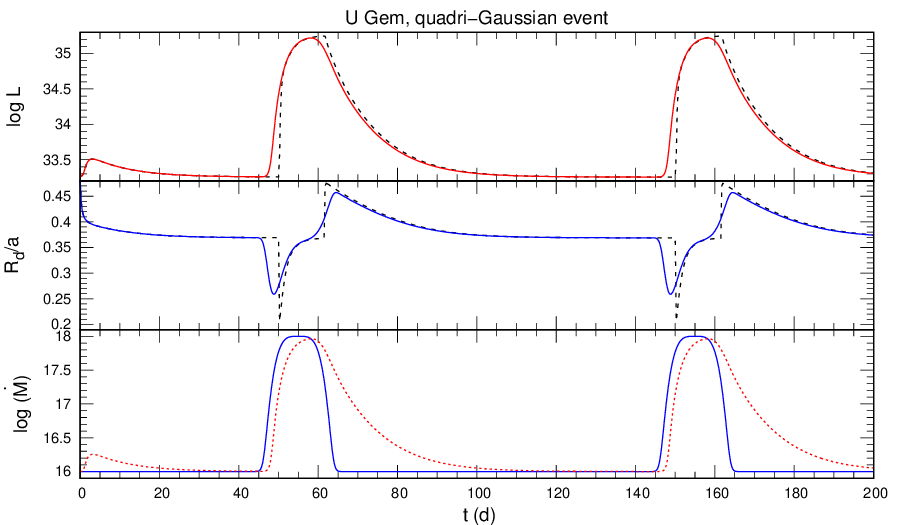}
\caption{Time evolution of the accretion disc for a quadri-Gaussian event. Analogous to Fig.~\ref{fig:teste_ichikawa_osaki1992}, but the dashed lines in the two upper panels are the results obtained in Sect.~\ref{sec:modelo}. \label{fig:pulso_gaussiano_4}}
\end{figure*}

Fig.~\ref{fig:pulso_gaussiano_4} shows the result of the simulation using the event shape of Eq.~\ref{eq:pulso} with $\Delta t_p = 10^6$\,s. We see that the total disc luminosity follows the mass transfer event behavior, with its asymmetric shape being driven by the viscous flow during the decline. For a high-viscosity disc ($\alpha \geq 1$), the shape of the outburst is highly correlated with the shape of the enhanced mass event. As for the radius of the disc, we now have a smoother transition. The disc shrinks during the rise to maximum, over a considerable fraction of the outburst, and reaches its largest radius during the decline, after the outburst maximum. The smooth progress of the event from Eq.~\ref{eq:pulso} leads to a delay between the increase in $\dot{M}$ in the outer regions of the disc and the response of the accretion onto the white dwarf (bottom panel), useful for explaining the well-known UV-delay effect observed in the initial phases of outbursts in several dwarf novae \citep{Schreiberet2005}. Finally, several dwarf novae exhibit a plateau during outburst maximum \citep[e.g., Z~Cam -][]{Warner1995}, and with this event shape the outburst maximum can be made rounded or with a plateau (depending on the width of $\Delta t_p$).

\subsection{Deposit of the gas stream}

Fig.~\ref{fig:teste_ichikawa_osaki1992} shows the results of the standard simulation of \citet{IchikawaOsaki1992}, where the mass is exclusively deposited at the outer edge of the disc, uniformly over a radial extent of $\Delta R = 0.1\,R_d$. Particularly noteworthy is the variation in the disc radius, shrinking at the beginning of the event but returning to the quiescent value while the event persists and expanding to its maximum value only at the end of the enhanced mass event phase. Since the lack of observations confirming the shrinkage of the disc at the beginning of the outburst was used as an argument to discard the MTIM, in this section we investigated the disc's response when we allow the gas stream to penetrate the disc and to deposit part of the mass in more internal disc regions. The stream penetration scenario was considered in the work of \citet{Bathet1983} and confirmed in 3D numerical simulations of the stream-disc interaction \citep{Bisikaloet1998a,Makitaet2000,Bisikalo2005}.

We investigated the effects on the outburst of different forms of how the gas stream mass is deposited in the accretion disc. First, we tested the influence of the shape of mass deposit, assuming that the stream material is deposited at the outer parts of the disc with a Gaussian shape centered at $R_d$ (i.e., only one side of the Gaussian is considered) instead of the box-like shape assumed by \citet{IchikawaOsaki1992}. A simulation using Gaussian deposit with a full-width-at-half-maximum of 0.1\,$R_d$ showed that the effect is negligible on luminosity and insignificant on the variation of $R_d$. Therefore, the shape of mass deposit does not influence the simulations.

Next, we considered the case where the radial extent of the gas stream mass deposit varies throughout the outburst. This is a plausible scenario because an increase in mass transfer rate proportionally increases the stream density and facilitates its penetration into the disc. For this purpose, we used the gas stream as modeled by \citet{Hessman1999} based on the tabulated data of \citet{LubowShu1975,LubowShu1976}. The expressions for $w_1$ and $w_2$ were revised by \citet{BaptistaSchlindwein2022} to take into account the updated $(2\pi/\gamma)^{1/2}$ values from the erratum of \citet{LubowShu2014}. The gas stream vertical ($H_s$) and horizontal ($W_s$) scale heights \citep[$\chi^{-1/2}$ and $\gamma^{-1/2}$, respectively, according to][]{LubowShu1975} are represented by separately fitted expressions,

\begin{eqnarray}
H_s (R,q,a) & \approx & \frac{1}{\sqrt{2\pi}} \, h_1(R/a) \, h_2(q) \, a\epsilon, \nonumber \\
W_s (R,q,a) & \approx & \frac{1}{\sqrt{2\pi}} \, w_1(R/a) \, w_2(q) \, a\epsilon,
\label{eq:deposito_variavel1}
\end{eqnarray}

\noindent where

\begin{eqnarray}
h_1 (R,a) & = & 0.060 + 3.17(R/a) - 2.90(R/a)^2, \nonumber \\
w_1 (R,a) & = & 0.389 + 2.21(R/a) - 2.46(R/a)^2, \nonumber \\
\log h_2 & \approx & +0.031 (\log q) + 0.095 (\log q)^2, \nonumber \\
\log w_2 & \approx & -0.023 - 0.067 (\log q) + 0.081 (\log q)^2, \nonumber \\
a\epsilon & = & \frac{c_{ss} \, P_\mathrm{orb}}{2\pi} = \frac{P_\mathrm{orb}}{2\pi} \, \left( \frac{k_b \, T_2}{\mu \, m_p} \right)^{1/2},
\end{eqnarray}

\noindent where $q = M_2/M_1$ is the binary mass ratio, $c_{ss}$ is the average isothermal sound speed at the $L_1$ point, $P_\mathrm{orb}$ is the orbital period, $k_b$ is the Boltzmann constant, $T_2$ is the effective temperature of the secondary, $\mu$ is the mean molecular weight, and $m_p$ is the proton mass. The factors $1/\sqrt{2\pi}$ in Eqs.~\ref{eq:deposito_variavel1} correct the mistake made by \citet{Hessman1999} by ignoring the additional factor $\sqrt{2\pi}$ included in the tables of \citet{LubowShu1975,LubowShu1976}.

The midplane density of the gas stream ($\rho_s$) is derived by integrating the stream density over all widths and heights and conserving mass \citep{Hessman1999},

\begin{equation}
\dot{M}_2 = 2\pi \, \rho_s \, W_s \, H_s \, V_s \,\,\, \longrightarrow \,\,\, \rho_s = \frac{\dot{M}_2}{2\pi \, W_s \, H_s \, V_s},
\label{eq:deposito_variavel2}
\end{equation}

\noindent where $V_s$ is the velocity of the gas stream,

\begin{equation}
V_s (R,q,a) = \sqrt{\frac{2 \, G \, M_1}{R_{L_1}}} \, v_s (R/R_{L_1}),
\end{equation}

\noindent and the relative velocity ($v_s$),

\begin{equation}
v_s (R/R_{L_1}) = 1.87 - 1.87 \frac{R}{R_{L_1}} + 4.1 \exp \left(\frac{-R}{0.085 \, R_{L_1}} \right). 
\end{equation}

Now, having an estimate of the midplane density of the gas stream (Eq.~\ref{eq:deposito_variavel2}), we can assume that the higher density stream will penetrate the lower density outer disc regions down to the radius $R_p$ where the midplane densities of the stream and the disc become equal. In the particular case of a steady thin disc, the midplane density is obtained by dividing the average density $\rho$ (Eq.~\ref{eq:conjunto_disco}) by $\sqrt{2\pi}$.

As we are interested in comparing the results of this simulation of the variable deposit with those obtained by \citet{IchikawaOsaki1992}, it is necessary to apply a scaling factor of 11 to Eq.~\ref{eq:deposito_variavel2} so that in quiescence the deposit occurs at $R \geq 0.9 \, R_d$. The mass is deposited uniformly between $R_d$ and the penetration radius of the stream ($R_p$).

\begin{figure*}[!htp]
\includegraphics[width=1.0\textwidth]{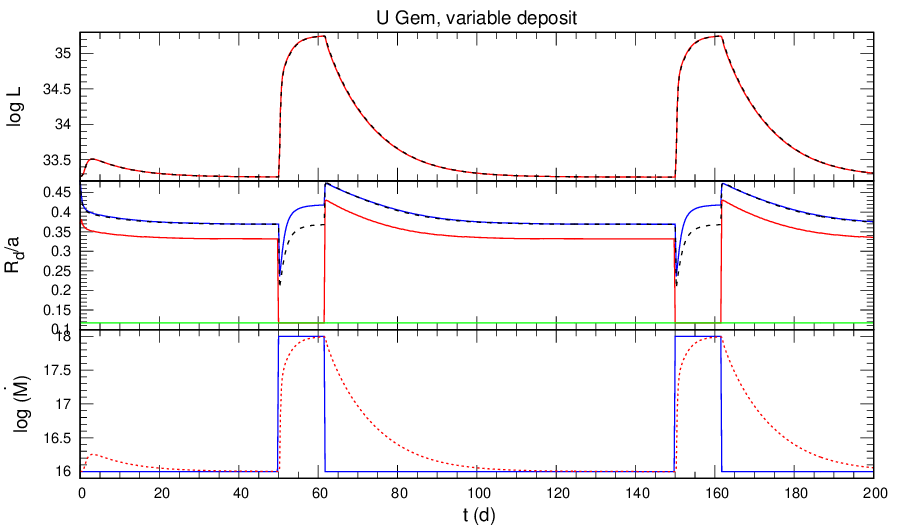}
\caption{Time evolution of the accretion disc for a variable deposit. Analogous to Fig.~\ref{fig:teste_ichikawa_osaki1992}, but the dashed lines in the two upper panels represent the results obtained in Sect.~\ref{sec:modelo}. In the middle panel, the green and red curves correspond to $R_c$ and $R_p$, respectively. \label{fig:deposito_variavel}}
\end{figure*}

Fig.~\ref{fig:deposito_variavel} shows the results obtained from a simulation using the mass deposit with variable radial extension described above. We notice that the total luminosity of the disc does not change with this mass deposit configuration. On the other hand, during the enhanced mass event, the disc radius in the simulation with variable deposit is significantly larger than that with deposit limited to the outer regions of the disc, expanding beyond the quiescent radius shortly after the transient reduction that occurs at the beginning of the mass event. This result is consistent with the increase in disc radius observed throughout the outburst in dwarf novae \cite[e.g.,][]{BaptistaCatalan2001}, and is in clear contrast to the increase in disc radius only at the beginning of the outburst decline resulting from simulations with deposit limited to the outer disc region. Taking into account the penetration of the disc by the gas stream (and the consequent extension of the mass deposit region in the disc) not only allows for a better description of the observed variations in disc radius throughout the outburst, but also enables inside-out outbursts previously unfeasible to reproduce with the MTIM. During the mass-transfer event, $R_p$ rapidly shrinks until it reaches $R_c$. This is because we assume $R_c$ as the smallest possible penetration radius of the stream into the disc for this simulation.

\subsection{Disc with gray atmosphere}

\citet{Tylenda1981} shows that the outer regions of accretion discs in CVs
with low mass-transfer rates ($\dot{M}_2 \lesssim 10^{16}\,g\,s^{-1}$) are
optically thin in the continuum, and that the optically thin regions expand
inwards as $\dot{M}_2$ decreases. Therefore, an additional refinement of our
simulation program is to go beyond the hypothesis of blackbody emission and
consider emission by a disc with a gray atmosphere, which allow for optically
thin disc regions.

We follow \citet{Tylenda1981} and assume that
(1) the disc radial structure is described by the standard $\alpha$-disc model
    \citep{ShakuraSunyaev1973};
(2) the disc is locally treated as a plane-parallel slab of matter uniform in
    the $z$ direction;
(3) the disc matter is in local thermodynamic equilibrium (LTE); and
(4) scattering of radiation is ignored.
From the standard disc model we have that the energy flux generated from surface
unit of the disc, $D(R)$, is given by,
\begin{equation}
D(R)= \frac{9}{8} \nu(R) \Sigma(R) \frac{G M_1}{R^3} \, .
\label{eq:dissipacao}
\end{equation}
With assumptions 3 and 4 the source function is simply the Planck function,
$S_\nu = B_\nu (T)$, and the formal solution of the radiative transfer equation
with assumption 2 gives the intensity of the radiation emitted in a direction
$\mu$ (where $\mu=\cos i$),
\begin{equation}
I_\nu = B_\nu (T) \left[ 1 - \exp \left( -\frac{\tau_\nu}{\mu} \right)\right] \, ,
 \label{eq:I_nu}
\end{equation}
where $\tau_\nu= \Sigma \, \kappa_\nu$ is the optical depth and $\kappa_\nu$ is
the absorption coefficient per mass. We calculate the radiative flux by
integrating the intensity over the solid angle,
\begin{equation}
F_\nu = \pi B_\nu (T) \left[ 1 - 2 \mathrm{E_3} (\tau_\nu) \right] \, ,
\end{equation}
where $\mathrm{E_3}$ is the integro-exponential function of the third order.
The energy conservation gives the equation,
\begin{equation}
D(R) = \int_0^{\infty} F_\nu (R) \; \text{d}\nu \, .
 \label{eq:D_F}
\end{equation}
The temperature is found from an iterative solution of Eqs.~\ref{eq:dissipacao}
and \ref{eq:D_F}, while the radiation spectrum of the entire disc observed at
an inclination angle $i$ is obtained by integrating Eq.~\ref{eq:I_nu} over the
disc surface. In the limit $\tau \gg 1$ (i.e., optically thick disc regions)
the solution falls back into the blackbody case, $D(R) = \sigma[T(R)]^4$,
because $\mathrm{E_3} (\tau) \cong 0$.

The calculation of the absorption coefficient $\kappa_\nu$ includes the
free-free (f-f) and bound-free (b-f) transitions of H$^0$, H$^-$, He$^0$ and
He$^+$, as well as H$_2^+$, He$_{ff}^-$, and electron scattering
\citep{Graybook}. Since the electron density depends on the temperature and
chemical composition, we assume solar composition in our calculations.

\begin{figure}[!htp]
\includegraphics[width=1.0\columnwidth]{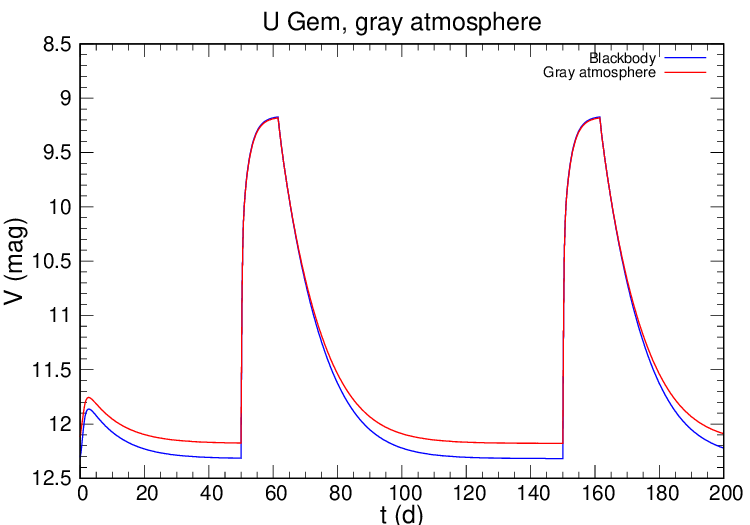}
\caption{Time evolution of the V magnitude of the accretion disc. The local emission from the disc is modeled using a blackbody (blue) or gray atmosphere (red). \label{fig:atmosfera_cinza_V}}
\end{figure}

We calculated the evolution of the magnitude in the V passband of the disc throughout the outburst by convolving the blackbody and gray atmosphere spectra with the V passband response curve \citep{Bessell1990}. The results can be seen in Fig.~\ref{fig:atmosfera_cinza_V}. The morphology of the light curve is the same in both cases, generating outbursts with an amplitude of $\sim 3$\,mag. The major difference occurs in quiescence, where the model with a gray atmosphere is brighter than the one with a blackbody. During the outburst, the magnitude of the two models is indistinguishable. This behavior can be explained by looking at Fig.~2 of \citet{Tylenda1981}. It is noted there that the higher the mass transfer rate, the more similar the spectrum of a gray atmosphere will be to that of a blackbody.

\begin{figure}[!htp]
\includegraphics[width=1.0\columnwidth]{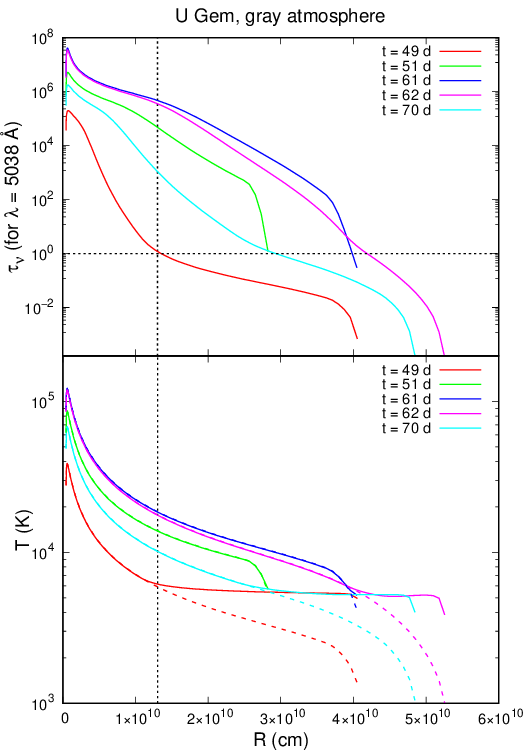}
\caption{Time evolution of optical depth (at 5038\,\AA, top
    panel) and effective temperature (bottom panel) for the simulations of
    Fig.~\ref{fig:atmosfera_cinza_V}. The horizontal dotted line in the top panel
    marks the $\tau=1$ limit below which the disc gas becomes optically thin.
    Temperatures in the bottom panel are computed using the blackbody (dashed
    lines) and gray atmosphere (solid lines) models for the disc emission.
    The notation is similar to that of Fig.~\ref{fig:teste_ichikawa_osaki1992_2}. \label{fig:atmosfera_cinza_teff}}
\end{figure}

Another interesting result concerns the behavior of the effective temperature.
The bottom panel of Fig.~\ref{fig:atmosfera_cinza_teff} shows the
evolution of $T_{\mathrm{eff}}$ during the outburst cycle for both models. As we can see, the $T_{\mathrm{eff}}$ of the gray atmosphere model becomes flat in the outer and optically thin regions of the disc, not falling below $\sim 6000$ K. Differences with respect to the $T_{\mathrm{eff}}\propto R^{-3/4}$ distribution of the blackbody emission model are more pronounced in quiescence, where the optically thin region of the disc extends down to $R_c$, than near the outburst maximum, where practically the entire disc is optically thick. The higher temperatures in the outer parts of the disc with a gray atmosphere explain the increased brightness of the quiescent disc with this model.

\subsection{Discussion}

The various tests presented in this section show that the variations in luminosity and outer radius of the disc over time can be modeled separately. Given a set of data to be modeled, we first adjust the time evolution of the total disc luminosity with the simple version of limited mass deposit, considering that the details of mass deposit on the disc do not affect the luminosity behavior. Subsequently, while fixing the best-fit luminosity, we then adjust the variation of $R_d$.

In summary, we can say that the phenomena that occur in radii smaller than the circularization radius affect the luminosity behavior, while those occurring at radii larger than $R_c$ affect the variation in $R_d$. This behavior led us to divide the disc ring grid into an inner region with exponential radial spacing (to better describe the regions with larger contribution to the disc luminosity) and an outer region with linear spacing (which helps in describing the variation of the disc radius).

The simulations presented in this work fill a few decades-long gap and allow the MTIM to be tested by comparison with a wide range of dwarf nova observations, with the same level of detail enjoyed by the DIM. It is worth mentioning that there are still points that can be explored, such as resolving the vertical structure of the disc, considering the magnetic moment of the primary and/or including the contribution/modeling of other light sources (white dwarf, secondary, and bright spot/gas stream), but these are beyond the scope of this paper.

\section{Application} \label{sec:aplicacao}

EX~Draconis (EX~Dra) is an eclipsing dwarf nova with an orbital period of 5.04\,h that shows outbursts with moderate amplitude ($\simeq 2$\,mag) and a recurrence timescale of $\simeq 20$-30\,d \citep{Baptistaet2000,Courtet2020}. Cyclical period changes on timescales of years, and a longer term trend of increasing orbital period been observed in the literature \citep{Baptistaet2000,Pilarciket2012,Courtet2020}. The distance inferred by Gaia~DR3 \citep{Gaia2016,GaiaDR3} is $241.2 \pm 1.3$\,pc, being consistent with the value measured by \citet{Baptistaet2000} of $290 \pm 80$\,pc through photometric parallax. The spectroscopy studies of \citet{Billingtonet1996,Fiedleret1997,SmithDhillon1998} led to a  `spectroscopic' model for the binary based on measurements of the radial velocity of the secondary ($K_2 \simeq 210-220$\,km\,s$^{-1}$) and the emission lines ($K_1 \simeq 163-176$\,km\,s$^{-1}$, associated with the orbital movement of the primary), and the rotational broadening of the secondary ($v \sin i = 140$\,km\,s$^{-1}$). \citet{Baptistaet2000} presented and discussed a set of light curves of EX~Dra in quiescence and outburst. The quiescent eclipse light curves were used to derive the geometry of the binary. The parameters obtained are listed in Table~\ref{tab:parametros_binarios}. \citet{ShafterHolland2003} did a multicolor photometric study of EX~Dra, where the binary parameters were also inferred. The photometric and spectroscopic models of the binary are consistent with each other within the uncertainties. The mass of the secondary suggests a mid-M type star, while the expected spectral type for this period would be near the K/M boundary. \citet{Harrisonet2004} derived a spectral type of K7, while \citet{Harrison2016} fitted the near-infrared spectrum of the secondary and obtained $T_2 = 4000$\,K. This value is consistent with what was modeled by \citet{ShafterHolland2003}.

\begin{deluxetable}{cc}[!htp]
\tablenum{1}
\tablecaption{The binary parameters of EX~Dra. \label{tab:parametros_binarios}}
\tablewidth{0pt}
\tablehead{}
%\decimalcolnumbers
\startdata
$q$                      & $0.72 \pm 0.06$          \\
$i (^\circ )$            & $85^{+3}_{-2}$           \\
$M_1/M_\odot$            & $0.75 \pm 0.15$          \\ 
$M_2/M_\odot$            & $0.54 \pm 0.10$          \\
$R_1/R_\odot$            & $0.011 \pm 0.002$        \\
$R_2/R_\odot$            & $0.57 \pm 0.04$          \\
$a/R_\odot$              & $1.61 \pm 0.10$          \\
$R_d/a$ (quies.)         & $0.267 \pm 0.004$        \\
$R_{L_1}/R_\odot$        & $0.85 \pm 0.04$          \\
$K_1$ (km s$^-1$)        & $163 \pm 11$             \\
$K_2$ (km s$^-1$)        & $224 \pm 17$             \\
$v_2 \, \sin i$ (km s$^-1$) & $136 \pm 9$              \\
\enddata
\tablecomments{Taken from \citet{Baptistaet2000}.}
\end{deluxetable}

\citet{BaptistaCatalan2001} analyzed the light curves from \citet{Baptistaet2000} with the eclipse mapping technique \citep{Horne1985,Baptista2016mapeamento}. Eclipse maps show evidence of the formation of a spiral arm in the disc in the early stages of the outburst and reveal how the disc expands during the rise phase until it fills most of the primary Roche lobe at maximum light. During the decay phase, the disc becomes progressively fainter until only a small bright region remains around the white dwarf at minimum light. Analysis of the radial brightness temperature distributions indicates that most of the disc appears to be in a steady state during quiescence and at outburst maximum, but not during the intermediate stages. As a general trend, the mass accretion rate in the outer regions is larger than that in the inner disc in the ascending branch, while the opposite is maintained during the descending branch. Fitting opaque steady disc models to radial temperature distributions allows one to estimate accretion rates of $\dot{M}= 10^{-7.7 \pm 0.3}$\,M$_\odot$\,yr$^{-1}$ ($1.6_{-0.8}^{+1.6} \times 10^{18}$\,g\,s$^{-1}$) at the outburst maximum and $\dot{M}= 10^{-9.1 \pm 0.3}$\,M$_\odot$\,yr$^{-1}$ ($5.0_{-2.5}^{+5.0} \times 10^{16}$\,g\,s$^{-1}$) in quiescence \citep{BaptistaCatalan2001}.

\begin{figure}[!htp]
\includegraphics[width=1.0\columnwidth]{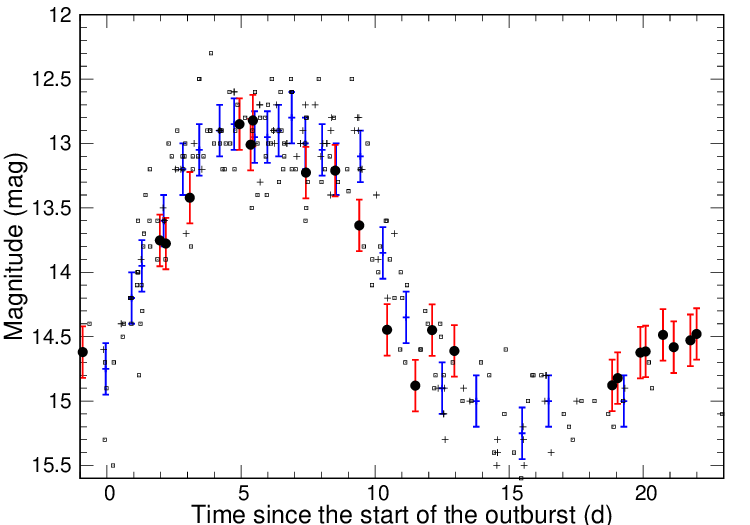}
\caption{Superposition of visual outburst light curves of EX~Dra, constructed from observations made by the AAVSO and the VSNET. The $x$-axis is time relative to the onset of the outburst. Crosses indicate measurements of the outbursts at the epoch of the observations of \citet{BaptistaCatalan2001}, while small dots are measurements of outbursts at other epochs. The blue points with error bars are the result of applying a median filter to this data set. R-band out-of-eclipse magnitudes from observations of \citet{Baptistaet2000} are shown as filled circles along with their uncertainties. \label{fig:curva_de_luz}}
\end{figure}

We used our MTIM outburst simulation code to model the variations in brightness and disc radius throughout the outburst of the dwarf nova EX~Dra. Fig.~\ref{fig:curva_de_luz} shows the historical visual light curve of EX~Dra (constructed from observations made by amateur astronomers from the AAVSO and VSNET) obtained from the superposition of observations covering 14 outburst cycles, aligned according to the start of the rise to maximum. The crosses indicate measurements of outbursts at the epoch of the observations of \citet{BaptistaCatalan2001}, while the dots are measurements of outbursts at other epochs. Only outbursts with amplitude and duration similar to those covered by the observations of \citet{BaptistaCatalan2001} were included. A median filter with a width of 10 points was applied to the data in this set (blue points with error bars). Filled circles mark the epochs of the observations of \citet{BaptistaCatalan2001} and indicate the corresponding R-band out-of-eclipse magnitudes. These are typical type B (inside-out) outbursts, with comparable rise and decline timescales \citep{Smak1984,Warner1995}. For the outbursts shown in Fig.~\ref{fig:curva_de_luz}, the rise from quiescence to maximum takes about 3\,d, followed by a plateau phase of about 6\,d. The declining branch lasts for 3-4\,d, after which the star apparently goes through a low-brightness state during 4-5\,d before recovering its quiescent brightness level \citep{BaptistaCatalan2001}.

We assume $\alpha$ constant, with a high value ($\geq 1$), and a variable deposit of the accretion stream on the disc. We allow the stream to penetrate the disc down to its shortest distance from the primary ($R_{min}$). This parameter is obtained from the computational trajectories of \citet{LubowShu1975} and can be approximated by the expression,

\begin{equation}
\frac{R_{min}}{a} = 0.0488 \, q^{-0.464},
\end{equation}

\noindent with an accuracy of 1\% in the range $0.05<q<1$. The radius of the shortest distance ($R_{min}$) is always smaller than the radius of circularization ($R_c$). The penetration radius of the stream ($R_p$) is given by the radius where the density of the gas stream equals the density of the midplane of the disc. The radius of the outer disc at each step along the outburst is defined as the corresponding tidal truncation radius. We adopted $\mu = 0.615$ (suitable for a solar abundance of fully ionized gases) and a tidal truncation constant of $c\omega= 30$\,rad/s \citep[Eq.~5 from][]{IchikawaOsaki1992} so that the quiescent disc radius coincided with the $R_d = (0.51 \pm 0.04) \, R_{L_1}$ value obtained by \citet{Baptistaet2000}. Since the EX~Dra outburst presents a plateau during maximum (Fig.~\ref{fig:curva_de_luz}), we adopted an enhanced mass event of the format described in Eq.~\ref{eq:pulso}. For the local accretion disc emission, we considered both the blackbody and gray atmosphere models. The resulting spectra were convolved with the V and R passband responses to generate the corresponding magnitude versus time curves. We adopted the binary parameters listed in Table~\ref{tab:parametros_binarios}.

\begin{figure}[!htp]
\includegraphics[width=1.0\columnwidth]{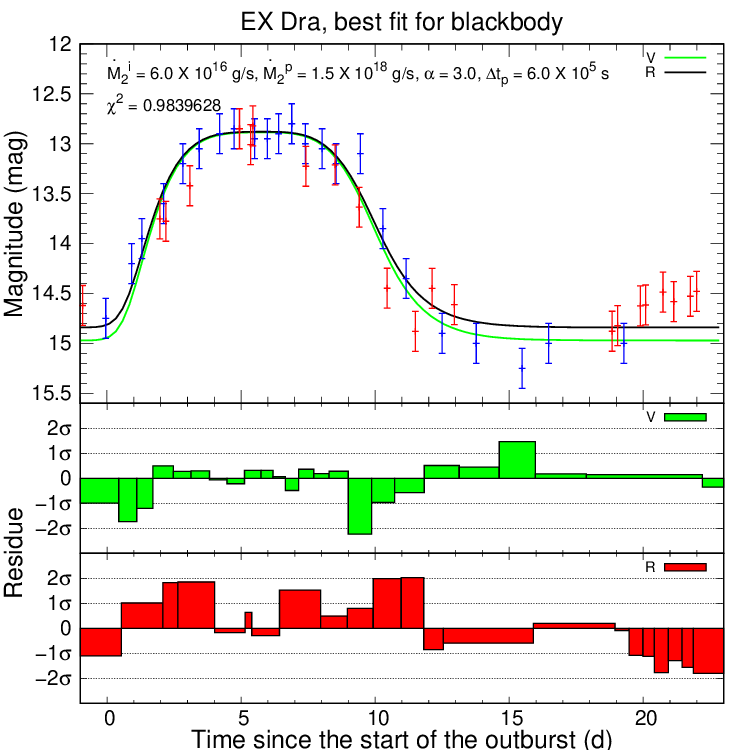}
\caption{Best-fit simulation in the case of local blackbody emission. The input parameters for this simulation are shown in the body of the figure together with the $\chi^2$ value of the fit. \textit{Upper panel:} The points with error bars are the same as in Fig.~\ref{fig:curva_de_luz}. The solid lines are the magnitudes in the V (green) and R (black) passbands of the simulated accretion disc. \textit{Bottom panels:} The residuals of each fit. \label{fig:ajuste_exdra_b}}
\end{figure}

\begin{figure}[!htp]
\includegraphics[width=1.0\columnwidth]{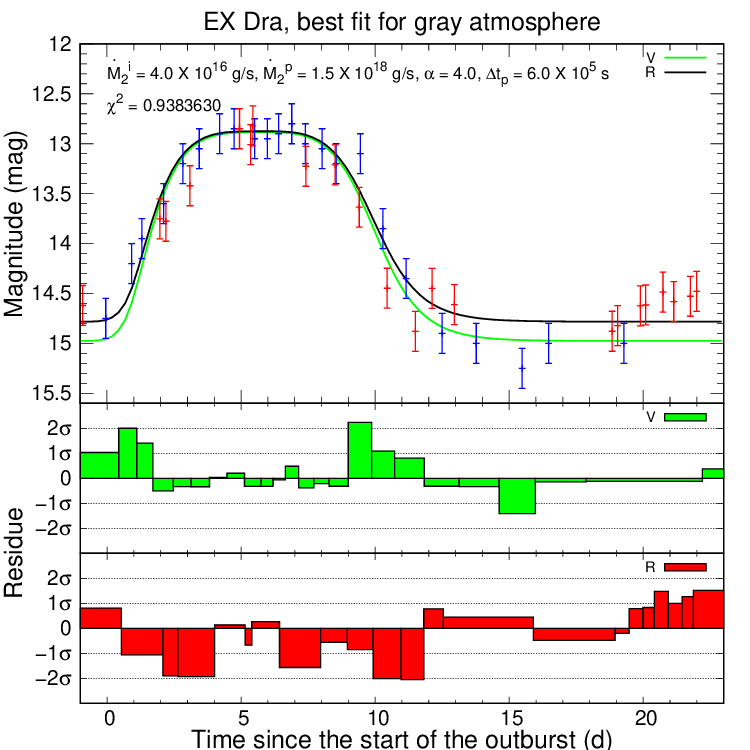}
\caption{Best fit simulation in the case of local emission by gray atmosphere. The notation is similar to that of Fig.~\ref{fig:ajuste_exdra_b}. \label{fig:ajuste_exdra_o}}
\end{figure}

We built a grid with 1000 outburst models, covering a range of values for the input parameters: quiescent mass transfer rate, $\dot{M}_2^i (= 2, 4, 6, 8, 10 \times 10^{16}$\,g/s), mass transfer rate during the maximum of the event, $\dot{M}_2^p (= 1.0, 1.5, 2.0, 2.5, 3.0 \times 10^{18}$\,g/s), event duration, $\Delta t_p (= 5.0, 5.5, 6.0, 6.5, 7.0 \times 10^5$\,s), and viscosity parameter, $\alpha (= 1, 2, 3, 4, 5, 6, 7, 8$). The value of $\dot{M}_2^i$ is determined by the magnitude in quiescence, and that of $\dot{M}_2^p$ is defined by the magnitude during the plateau. The value of $\Delta t_p$ is determined by the full-width-at-half-maximum of the outburst and the duration of the plateau. The rise timescale is determined by the event shape and the value of $\Delta t_p$. The value of $\alpha$ is determined by the outburst decline timescale. The best fit outburst model to the observations was found by calculating the $\chi^2$ of the fit for each grid model. Figs.~\ref{fig:ajuste_exdra_b} and \ref{fig:ajuste_exdra_o} show the best-fit simulations for the cases of blackbody and gray atmosphere local emission, respectively.

The MTIM model with a mass-transfer event given by Eq.~\ref{eq:pulso} and high viscosity provides a satisfactory description of the observations of the outbursts in EX~Dra, with $\chi^2$ values close to unity. The best-fit $\dot{M}_2^i$ and $\dot{M}_2^p$ values are consistent with the quiescence and outburst accretion rates inferred from the radial brightness temperature distributions of \citet{BaptistaCatalan2001} within the respective uncertainties. The inferred $\alpha$ values are high, but consistent with the range of values inferred by \citet{MantleBath1983} from the outburst decline timescale in dwarf novae. Furthermore, the amplitude of the outburst in the V passband is larger than that in the R passband, in accordance with observations. To our knowledge, this is the first time that a simulation of dwarf nova outbursts has been directly compared with observational data using a $\chi^2$ test. Typically, data and simulations are plotted in different figures in the literature, as DIM simulations do not describe the dwarf novae quiescence phase well.

\begin{figure}[!htp]
\includegraphics[width=1.0\columnwidth]{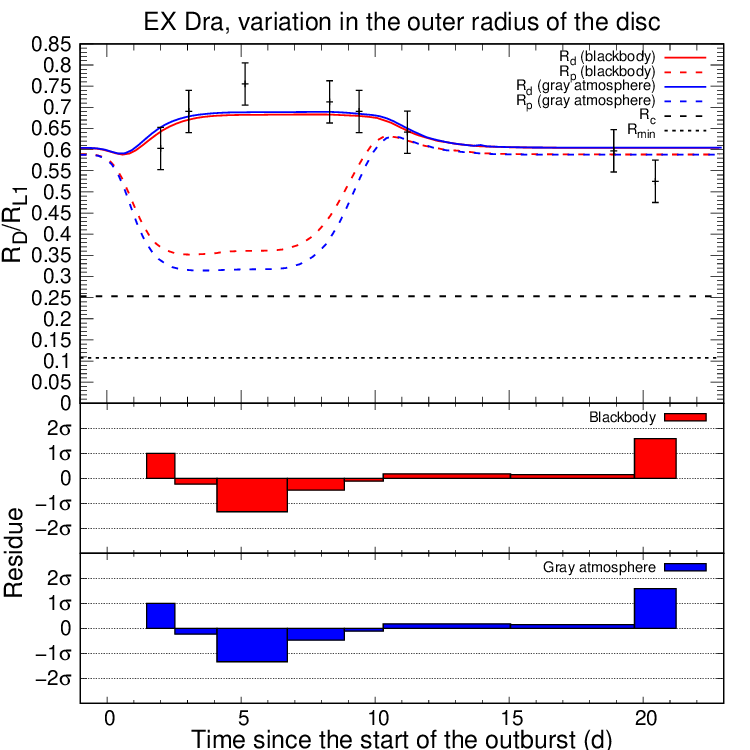}
\caption{Comparison of the variation in the outer disc radius from our simulations with the observational results of \citet{BaptistaCatalan2001}. \textit{Upper panel:} The points with error bars are the measurements of \citet{BaptistaCatalan2001}, while the red and blue lines represent the case of local emission by blackbody and gray atmosphere, respectively -- with the solid line being the outer disc radius and the dashed line being the radius of stream penetration. The black dashed and dotted lines mark the value of the circularization radius and the minimum radius of \citet{LubowShu1975}, respectively. \textit{Bottom panels:} The residuals of each fit. \label{fig:ajuste_exdra_raio}}
\end{figure}

Fig.~\ref{fig:ajuste_exdra_raio} compares the observed evolution of the outer disc radius as measured by \citet{BaptistaCatalan2001} with the prediction of the best-fit models in Figs.~\ref{fig:ajuste_exdra_b} and \ref{fig:ajuste_exdra_o}. It should be noted that, while \citet{BaptistaCatalan2001} estimated the disc radius from the point in the radial distribution where the intensity is equal to the maximum intensity of the bright spot in quiescence, the simulations estimate the value of $R_d$ from the radius of truncation of the accretion disc. These definitions are different and, therefore, may lead to different results throughout the outburst. Despite this, the best-fit models provide a good description of the observed variations in outer disc radius throughout the outburst, with the largest difference remaining below the 2-$\sigma$ level.

Our models predict an almost imperceptible shrinking of $R_d$ at the beginning of the outburst followed by an immediate increase in $R_d$, with the maximum radius being reached just before maximum brightness. The existence of a gas stream penetration into the disc eliminates the marked initial reduction in $R_d$ present in the \citet{LivioVerbunt1988} and \citet{IchikawaOsaki1992} simulations and leads to an outer disc radius variation consistent with the observations. The amplitude of the $R_d$ variation in our simulations is smaller than observed; the MTIM models tend to underestimate the outburst maximum disc radius and to overestimate the quiescent disc radius. This may be due to differences between the definitions of $R_d$ in \citet{BaptistaCatalan2001} and that of the simulation program, or to gas stream mass being deposited in the accretion disc in a different way than assumed in the simulations. In this latter case, a more realistic treatment requires an analytical description of the interaction between the stream and disc particles based on 3D simulations of the accretion stream. There is still no work in this regard in the literature. Alternatively, \citet{BuatMenardet2001} suggest that it is possible to reproduce larger amplitude variations in the disc radius throughout the outburst by reducing the exponent of the radial dependence of the term of the tidal effect.

\section{Discussion} \label{sec:discursao}

\subsection{Is $\alpha>1$ unphysical?}

In their effort to overcome the absense of a theory of viscosity in differentially rotating fluid discs \citep{ShakuraSunyaev1973,ShakuraSunyaev1976} parametrized the kinematic viscosity in terms of the known quantities $c_s$ and $H$, and transferred the ignorance on the unknown viscosity mechanism to the non-dymensional $\alpha$ parameter,
\begin{equation}
\nu = \alpha c_s H \, ,
\end{equation}
where
\begin{equation}
\alpha = \frac{v_t l_t}{c_s H} + \frac{B^2}{4\pi \rho c_s^2} \, ,
\end{equation}
$v_t$ and $l_t$ are respectively the turbulent velocity and the turbulent mixing length (in case of hydrodynamic turbulence), $B^2/8\pi$ is the energy density of the chaotic magnetic field and $\rho c_s^2/2$ is the thermal energy density of the matter in the disc \citep[in case of magneto-hydrodynamic turbulence, e.g.,][]{BalbusHawley1991,HawleyBalbus1991}. The theoretical expectation $\alpha \lesssim 1$ is based on the assumptions that (i) the size of the largests turbulent eddies can not exceed the disc thickness ($l_t \leq H$) and (ii) the turbulence must be subsonic ($v_t \lesssim c_s$) otherwise the turbulent motions would probably be termalized by shocks \citep{ShakuraSunyaev1973,Franket2002}.

\citet{Franket2002} noted that $v_t > c_s$ could occur in disc regions where some physical input continually feeds a supersonic turbulence. \citet{Martinetal2019} remarked that both the disc mass inflow and the rate of heating by viscous dissipation depend linearly on the value of $\alpha$ and that, while a larger $\alpha$ leads to a larger amount of energy dissipation (presumably through shocks), it also leads to a larger accretion rate which can provide the necessary larger energy to be dissipated. Hence, they concluded that the $v_t \lesssim c_s$ assumption is not justified on energy arguments. In addition, we argue that there is no reason to exclude the possibility that shocks in trans-sonic or supersonic turbulence may just provide the dissipative process required for viscous heating in accretion discs.

The $l_t \leq H$ assumption is based on the idea that turbulence is isotropic. However, in an accretion disc angular momentum exchange occur in the radial direction, strong stretching of turbulent eddies or magnetic field lines occur in the azimuthal direction, and most of the energy resulting from viscous dissipation is transported in the vertical direction. It makes not much sense to believe that (hydrodynamical or magneto-hydrodynamical) turbulence in such an anisotropic environment should be isotropic. Indeed, numerical simulations of the magneto-rotational instability show how an initially isotropic magnetic field rapidly becomes chaotic, anisotropic and significantly stretched in the radial direction, with the amount of field stretching being limited mostly by the radial width of the shearing box \citep[][e.g., see their Fig.~9]{HawleyBalbus1991}. In addition, the radial stretching of the magnetic field lines allows exchange of angular momentum over radial distances which greatly exceed the linear wavelength of the instability (which will be present as long as the minimum unstable wavelength do not exceed the disc thickness), thus indicating that $l_t > H$.

Given that both $l_t > H$ and $v_t > c_s$ are valid possibilities, the theoretical expectation $\alpha \lesssim 1$ seems unjustified and there is no {\em a priori} reason to consider that $\alpha > 1$ is unphysical.

\subsection{DIM x MTIM: The EX~Dra case}

Observations of eclipsing dwarf novae throughout their ourburst cycle provide a key opportunity to test the predictions and to discriminate between the DIM and MTIM proposed explanations \citep[e.g.,][]{Baptista2012}. In particular, the critical tests are those the results of which are consistent with predictions of one of the models, but inconsistent with those of the other.

For EX~Dra, \citet{BaptistaCatalan2001} found that the inward-moving cooling wave during decline decelerates as it travels inwards, in line with DIM predictions \citep{Menouet1999}. In the MTIM framework, the radial drift velocity $v_R$ may also decrease inwards if the viscosity parameter increases with radius, as expected when gas stream-outer disc interactions and tidal dissipation effects are important (see Sect.~\ref{sec:alpha_variavel}). Accordingly, the higher inwards velocity ($v_R= -0.75 \pm 0.26$\,km\,s$^{-1}$) corresponds to the early outburst decline phase where the outer disc radius is still larger than the tidal truncation radius \citep[$R_d \simeq 0.7\,R_{L_1} > R_\mathrm{tid}=0.66\,R_{L_1}$][]{Paczynski1977}, the cooling wave is moving through the outer disc regions and there is still significant gas stream penetration occurring (e.g., Fig.~\ref{fig:ajuste_exdra_raio}), while the lower inwards velocity ($v_R= -0.46 \pm 0.23$\,km\,s$^{-1}$) corresponds to the later  decline phase where the outer disc radius shrinks below the tidal truncation radius, the cooling wave is moving through intermediate disc regions and the MTIM simulations predict negligible gas stream penetration. Since this result is consistent with both DIM and MTIM, it offers no power to discriminate between these models \citep{Baptista2016mapeamento}.

On the other hand, DIM predicts that quiescent dwarf novae harbor low-viscosity \citep[$\alpha_\mathrm{cool} \leq 0.05$, e.g.,][]{Hameuryet1998}, non-steady state accretion discs with a flat radial temperature distribution and a slow viscous response to eventual changes in mass transfer rate, while MTIM predicts high-viscosity steady-state quiescent discs, with a fast response to changes in mass transfer rate. The differences in prediction are very significant. For EX~Dra ($R_d= 3\times 10^{10}$\,cm) the estimated viscous timescale of a quiescent DIM disc is (cf. Eq.~\ref{eq:alpha_DIM_MTIM}),

\begin{equation}
  t_\mathrm{visc}(\mathrm{DIM}) \simeq \frac{R_d}{12\,\alpha_\mathrm{cool}\, c_s}
  \frac{R}{H} \simeq 8.3 \, \frac{R_d}{\alpha_\mathrm{cool}\, c_s}
  \geq 60\,d  ,
\end{equation}

\noindent which is much longer than its $(20-30)\,d$ outburst recurrence time -- indicating that there is not enough time to reach a state-state during quiescence. However, the eclipse mapping analysis of \citet{BaptistaCatalan2001} reveals that both the radial temperature distribution at minimum light (their eclipse map $g$) and in quiescence (their eclipse map $h$) closely follow the $T\propto R^{-3/4}$ law of opaque steady-state discs -- respectively with mass accretion rates of $\dot{M}(g)= (2.3\pm 0.3)\times 10^{16}$\,g\,s$^{-1}$ and $\dot{M}(h)= (5.3\pm 0.6)\times 10^{16}$\,g\,s$^{-1}$ -- and that the transition between these two steady-states occur in a viscous timescale of only $t_\mathrm{visc}(\mathrm{quies})= 1.5\,d$, implying a viscosity parameter of,

\begin{equation}
  \alpha_\mathrm{quies} \simeq 8.3
  \frac{R_d}{c_s\,t_\mathrm{visc}(\mathrm{quies})} \simeq 2 ,
\end{equation}

\noindent in line with the high viscosity range inferred in Sect.~\ref{sec:aplicacao}. These results are in clear contradiction with DIM and in agreement with the MTIM expectations.

Moreover, the early rise map \citep[eclipse map $b$ of][]{BaptistaCatalan2001} shows evidence of enhanced gas stream emission, indicative of an enhanced mass transfer rate at this early outburst stage. The integrated disc luminosity at early rise ($1.4 \pm 0.3\,L_\odot$) is comparable to that in quiescence ($1.2 \pm 0.1\,L_\odot$) and is not enough to support the idea that the enhanced mass transfer could be triggered by an increased irradiation of the secondary star by the accretion disc. This led \citet{Baptista2012} to the conclusion that the observed enhanced mass transfer at early rise is not a consequence of the ongoing outburst, but its cause -- suggesting that the outbursts of EX~Dra are powered by bursts of enhanced mass transfer, as expected by MTIM.

The scenario which emerges from these previous EX~Dra results clearly favors MTIM. This is underscored by the results from this paper, which shows that the observed variations in brightness and outer disc radius of EX~Dra throughout its outburst cycle can be described, with a unity chi-square fit, by an MTIM outburst model of a $6\times 10^5$\,s long, 30-fold increase in mass transfer rate event onto a high-viscosity $\alpha= 3-4$ standard accretion disc with significant gas stream penetration.

\section{Conclusion} \label{sec:conclusao}

In this work, we present the results of the development of a simulation program for the response of accretion discs to events of enhanced mass transfer, in the context of MTIM. Our program resumes simulations with MTIM based on the work of \citet{IchikawaOsaki1992}, the last publication on the subject found in the literature. The developed program was validated by reproducing the results obtained by \citet{Pringle1981} and \citet{IchikawaOsaki1992}.

We expanded the horizon of the MTIM simulations by investigating the effects of (i) a radial dependency on the viscosity parameter, (ii) a smoother enhanced mass-transfer event profile, (iii) mass deposition with gas stream penetration, and (iv) disc emission with gray atmosphere. An increasing $\alpha$ with radius (e.g., due to additional energy dissipation in the outer disc regions resulting from stream-disc interaction and the tidal influence of the secondary star) helps to reduce the long exponential outburst decay of previous simulations. $\alpha \geq 1$ are required in order to match the observed outburst decline timescales of dwarf novae, as previously found by  \citet{MantleBath1983}. A more realistic, smoother enhanced mass-transfer event profile results in continuous changes in disc radius along the outburst and lead to a delay between the increase in $\dot{M}$ in the outer regions and the accretion onto the white dwarf, akin to the well known UV-delay effect seen in the initial outburst stages of several dwarf novae. Gas stream penetration allows for inside-out outbursts with minimal disc shrinkage at outburst onset. Gray atmospheres allow for optically thin outer disc regions and lead to correspondingly brighter quiescent discs.

We applied our MTIM simulation code to the EX~Dra data of \citet{BaptistaCatalan2001}. The observed variations in brightness and outer disc radius throughout the inside-out outbursts of EX~Dra are well described ($\chi^2\simeq 1$) by the response of a high-viscosity accretion disc to an event of enhanced mass transfer rate, in accordance with MTIM predictions. The best-fit models indicate $\alpha= 3-4$ and a mass transfer event of width $\Delta t= 6\times 10^5$\,s. The values of the quiescence and outburst maximum mass transfer rates inferred from the MTIM simulations are consistent with those derived by \citet{BaptistaCatalan2001} at the 1-$\sigma$ limit. Our results underscore the previous suggestion by \citet{Baptista2012} that the outbursts of EX~Dra are powered by burst of enhanced mass transfer.

It is a good scientific practice to let the observations determine the range of possible $\alpha$ values, and to put aside the theoretical expectation if it proves inconsistent with observations. In dwarf novae, the observed outburst decline timescales lead to $\alpha_\mathrm{MTIM} \simeq 10\,\alpha_\mathrm{DIM} \simeq 1-3$ \citep[for hydrogen-rich discs, e.g.,][]{MantleBath1983,Bathet1983,Cannizzo2001,Colemanetal2016} and $\alpha_\mathrm{MTIM} \simeq 3-4$ \citep[for the hydrogen-deficient disk of YZ\,LMi,][]{BaptistaSchlindwein2022}. Moreover, modeling of soft x-ray transient (SXT) outburst light curves lead to $\alpha_\mathrm{MTIM} \simeq 10\,\alpha_\mathrm{DIM} \approx 2-10$ \citep{Dubusetal2001,Tetarenkoetal2018}. Therefore, if dwarf nova and SXT outbursts are viscous events (instead of thermal-viscous instability events), than the inferred viscosity parameters are systematically larger than unity and the theoretical expectation $\alpha \lesssim 1$ is inconsistent with observations.

%% IMPORTANT! The old "\acknowledgment" command has be depreciated. It was
%% not robust enough to handle our new dual anonymous review requirements and
%% thus been replaced with the acknowledgment environment. If you try to 
%% compile with \acknowledgment you will get an error print to the screen
%% and in the compiled pdf.
\begin{acknowledgments}
We thank the anonymous referee for useful comments and suggestions that helped to improve the presentation of our results. This study was financed in part by the Coordenação de Aperfeiçoamento de Pessoal de Nível Superior - Brasil (CAPES) - Finance Code 001. WS acknowledges financial support from CNPq/Brazil (Proc. 301366/2023-3, 300252/2024-2 and 301472/2024-6). WS is grateful for the discussions on numerical methods with Leandro Roza Livramento.
\end{acknowledgments}

%% To help institutions obtain information on the effectiveness of their 
%% telescopes the AAS Journals has created a group of keywords for telescope 
%% facilities.
%
%% Following the acknowledgments section, use the following syntax and the
%% \facility{} or \facilities{} macros to list the keywords of facilities used 
%% in the research for the paper.  Each keyword is check against the master 
%% list during copy editing.  Individual instruments can be provided in 
%% parentheses, after the keyword, but they are not verified.

\vspace{5mm}
\facilities{0.9\,m James Gregory Telescope, AAVSO, VSNET}

%% Similar to \facility{}, there is the optional \software command to allow 
%% authors a place to specify which programs were used during the creation of 
%% the manuscript. Authors should list each code and include either a
%% citation or url to the code inside ()s when available.

%\software{ASTROPOP \citep{Campagnolo2019}}

%% Appendix material should be preceded with a single \appendix command.
%% There should be a \section command for each appendix. Mark appendix
%% subsections with the same markup you use in the main body of the paper.

%% Each Appendix (indicated with \section) will be lettered A, B, C, etc.
%% The equation counter will reset when it encounters the \appendix
%% command and will number appendix equations (A1), (A2), etc. The
%% Figure and Table counter will not reset.

\appendix

\section{Steady thin disc} \label{sec:apendice_disco}

In revising the set of equations for a steady thin disc using the prescription of \citet{ShakuraSunyaev1973}, we noticed differences between our expressions and those listed by \citet{Franket2002}. The main reason for these differences is the change in the expression of the Rosseland mean opacity ($\kappa_R$) between the second and third edition of the \citet{Franket2002} book. While those authors state that the change only affects the expression of the optical depth ($\tau$), we found that other equations are also modified by the change in $\kappa_R$. Below we list the corrected set of equations used in our MTIM simulation code:

\begin{equation}
\left.
\begin{array}{rl}
\Sigma &= 5.30 \alpha^{-4/5} \mu^{3/4} \dot{M_{16}}^{7/10} m_1^{1/4} R_{10}^{-3/4} f^{14/5} \; \text{g cm}^{-2}, \\
H &= 1.32\times 10^8 \alpha^{-1/10} \mu^{-3/8} \dot{M_{16}}^{3/20} m_1^{-3/8} R_{10}^{9/8} f^{3/5} \; \text{cm}, \\
\rho &= 4.02\times 10^{-8} \alpha^{-7/10} \mu^{9/8} \dot{M_{16}}^{11/20} m_1^{5/8} R_{10}^{-15/8} f^{11/5} \; \text{g cm}^{-3}, \\
T_c &= 2.80\times 10^4 \alpha^{-1/5} \mu^{1/4} \dot{M_{16}}^{3/10} m_1^{1/4} R_{10}^{-3/4} f^{6/5} \; \text{K}, \\
\tau &= 291 \alpha^{-4/5} \mu \dot{M_{16}}^{1/5} f^{4/5}, \\
\nu &= 2.00\times 10^{14} \alpha^{4/5} \mu^{-3/4} \dot{M_{16}}^{3/10} m_1^{-1/4} R_{10}^{3/4} f^{6/5} \; \text{cm}^2 \text{ s}^{-1}, \\
v_R &= -3.00\times 10^4 \alpha^{4/5} \mu^{-3/4} \dot{M_{16}}^{3/10} m_1^{-1/4} R_{10}^{-1/4} f^{-14/5} \; \text{cm s}^{-1} \\
\text{with } f &= \left[ 1 - \left(\frac{R_1}{R} \right)^{1/2} \right]^{1/4},
\end{array}
\right\}
\label{eq:conjunto_disco}
\end{equation}

\noindent where $\Sigma$ is the surface density, $H$ is the scaleheight, $\rho$ is the density, $T_c$ is the central temperature of the disc, $\tau$ is the optical depth, $\nu$ is the viscosity, $v_R$ is the radial drift velocity, $\mu$ is the mean molecular weight, $\dot{M}_{16} = \dot{M_2}/(10 ^{16} \text{ g s}^{-1})$ with $\dot{M_2}$ being the mass transfer rate, $m_1 = M_1/M_\odot$ with $M_1$ being the mass of the primary, $ R_{10} = R/(10^{10} \text{ cm})$ where $R$ is the radial coordinate, and $R_1$ is the radius of the primary. 

%% For this sample we use BibTeX plus aasjournals.bst to generate the
%% the bibliography. The sample631.bib file was populated from ADS. To
%% get the citations to show in the compiled file do the following:
%%
%% pdflatex sample631.tex
%% bibtext sample631
%% pdflatex sample631.tex
%% pdflatex sample631.tex

\bibliography{ref}{}
\bibliographystyle{aasjournal}

%% This command is needed to show the entire author+affiliation list when
%% the collaboration and author truncation commands are used.  It has to
%% go at the end of the manuscript.
%\allauthors

%% Include this line if you are using the \added, \replaced, \deleted
%% commands to see a summary list of all changes at the end of the article.
%\listofchanges

\end{document}